\documentclass[useAMS,usenatbib]{mn2e}
\usepackage{graphicx}

\title[Spectral states of EXO 1745--248]{Hysteresis in the spectral states 
of the neutron star low-mass X-ray binary EXO 1745--248}
\author[Mukherjee and Bhattacharyya]{Arunava Mukherjee$^{1}$\thanks{E-mail:
arunava@tifr.res.in} and Sudip Bhattacharyya$^{1}$\thanks{E-mail:
sudip@tifr.res.in} \\
$^{1}$Department of Astronomy and Astrophysics, Tata Institute
of Fundamental Research, Mumbai 400005, India}

\begin{document}

\date{
}

\pagerange{\pageref{firstpage}--\pageref{lastpage}} \pubyear{2010}

\maketitle

\label{firstpage}
\begin{abstract}
We study the low-frequency timing properties and the spectral state
evolution of the transient neutron star low-mass X-ray binary
EXO 1745--248 using the entire {\it Rossi X-ray Timing Explorer}
Proportional Counter Array data. We tentatively conclude that EXO 1745--248 
is an atoll source, and report the discovery of a $\approx 0.45$ Hz 
low-frequency quasi-periodic oscillation and $\sim 10$ Hz peaked noises.
If it is an atoll, this source is unusual because (1) instead of a `C'-like curve,
it traced a clear overall clockwise hysteresis curve in each of the colour-colour
diagram and the hardness-intensity diagram; and (2) the source
took at least 2.5 months to trace the softer banana state,
as opposed to a few hours to a day, which is typical for an atoll source. 
The shape of the hysteresis track was intermediate between 
the characteristic `q'-like curves of several black hole systems and 
`C'-like curves of atolls, implying that EXO 1745--248 is an important source
for the unification of the black hole and neutron star accretion processes.
\end{abstract}


\section{Introduction}\label{Introduction}

The spectral states and the correlated timing properties of neutron star and 
black hole low-mass X-ray binaries (LMXBs) can be very useful to understand 
the extreme environments of these sources \citep{vanderKlis2006}. An excellent 
way to study these properties is to track these sources in the colour-colour
diagram (CD; hard colour (HC) vs. soft colour (SC))
and in the hardness-intensity diagram (HID; hard colour vs. intensity; 
see \S~\ref{DataAnalysisandResults}). 
From the beginning of an outburst, the intensity of a transient black hole 
source increases, typically keeping the HC at a near-constant value.
Near the highest intensity the HC value quickly decreases, followed by
an intensity decrease at a lower HC value, and a soft-to-hard transition at
a lower intensity value. Thus a black hole LMXB typically traces a `q'-like
hysteresis curve in the HID \citep{vanderKlis2006, Belloni2009}.
It is usually believed that neutron star LMXBs do not trace hysteresis curves
in CD/HIDs \citep{vanderKlis2006}. For example, the near-Eddington Z sources trace out roughly `Z' shaped
tracks on time scales of hours to a day, while the less luminous atoll sources
have `C' shaped tracks \citep{vanStraatenetal2003, vanderKlis2006}. 
The lower HC banana-like portion (BS) of the `C' track can be divided into 
upper banana (UB), lower banana (LB) and lower left banana (LLB) based on
spectral and timing properties. The BS is traced out on time scales of hours to a day 
without any hysteresis \citep{vanderKlis2006}. On the other hand, 
the higher HC extreme island state (EIS) is traced out in days to weeks, and
secular motions in the form of parallel tracks are seen in EIS. 
An atoll source moves from EIS to BS via an island state (IS).
Probably the only transient atoll source showing a `q'-like hysteresis HID curve is Aql X-1
(\citet{MaitraBailyn2004, Reigetal2004}; see also \citet{Bellonietal2007} 
for 4U 1636--53 tracks). Such neutron star LMXBs, and more importantly
sources showing intermediate tracks between `q' and `C', can be very useful (1) to unify the
black hole and neutron star accretion processes, and (2) to sort out the 
mismatch between the standard EIS-IS-BS
framework and the general hysteresis phenomena.
In this Letter, we show that the bursting neutron star LMXB EXO 1745--248
\citep{MarkwardtSwank2000, Wijnandsetal2002, Heinkeetal2003}
is such an intermediate source with unique properties.

\section{Data Analysis and Results}\label{DataAnalysisandResults}

The neutron star transient LMXB EXO 1745--248 was observed with 
{\it Rossi X-ray Timing Explorer} ({\it RXTE}) in 2000 and 2002: (1) between Jul 13, 
2000 (start time: 04:43:28) and Nov 3, 2000 (end time: 00:17:52; proposal nos.: P50054 and P50138); and (2) between Jul 2, 2002 (start time: 20:38:24) and Jul 22, 2002 (end time: 11:04:00; proposal no.: P70412)
for a total observation time of 144 ks. 
We have produced CD and HID using the entire standard-2 mode data from the top layers of Proportional Counter Unit (PCU) 2. We have defined HC and SC as the ratio of the background-subtracted detector counts in the $(9.2-18.9)/(5.7-9.2)$ and $(3.9-5.7)/(2.6-3.9)$ keV energy bands, respectively.
We have been able to divide the 2000 data in nine temporal segments, i.e., nine phases (see Table 1 for time ranges). Each phase traces a distiguishable portion of the HID track (see Fig.~\ref{HID}). This figure shows that the source starts from a low intensity and a high HC value, and in the hard state (phase 1--4) it traces a few adjacent curved parallel tracks below the intensity $\sim 450$ counts/s/PCU and in the HC range of $\sim 0.6-1.3$. Unlike Aql X--1 and several black-hole sources, the intensity of EXO 1745--248 does not increase much in the highest HC value. Rather, Fig.~\ref{HID} shows that the HC value decreases substantially and the intensity increases at a lower HC value of $\sim 0.6$ in the hard state. During the transition from phase 4 to phase 5, the source goes through a large intensity (in $2.6-18.9$ keV) jump from $\sim 361$ counts/s/PCU to $\sim 1135$ counts/s/PCU while having a relatively small change in the HC value ($\sim 0.60$ to $\sim 0.36$; see Table 1 and Fig.~\ref{HID}). In the high intensity state (phase 5--7) the source shows a clear anti-clockwise loop (hysteresis; Fig.~\ref{HID}). EXO 1745--248 undergoes a moderate intensity jump from phase 7 to phase 8, and the intensity steadily decreases up to phase 9 while keeping the HC value nearly same (Fig.~\ref{HID}). The lower intensity portion of phase 9 shows a significant increase in HC value. Since the ASM data confirm that the source intensity further decreases into the quiescence, Fig.~\ref{HID} implies a clear overall clockwise loop (hysteresis) of the source. The Intensity and HC values of phase 10 (2002 data) are consistent with those of phases 8 and 9. The phases 5--10 show a clear banana-like track in the CD (Fig.~\ref{CCD}). In the hard state, two phases (1,3), which display substantial changes in HC values, show large variation in SC values.

In order to identify the spectral states of EXO 1745--248, we have analyzed the low-frequency power spectra of each phase using all the PCA event-mode data with a standard technique \citep{vanderKlis1989}. Each Leahy-normalized power spectrum has a Nyquist frequency of 128 Hz and the best resolution of 0.004 Hz. We have fitted the continuum component of a power spectrum with a constant+powerlaw model (describing white and red noises, respectively) and any narrow feature with a Lorentzian. The hard state (Phase 1--4) power spectra are typically well described with a constant+powerlaw, having a strong red noise (typical RMS 
$\sim 25 - 45 \%$) below $\sim 10$ Hz with the Leahy-power reaching $> 1000$ at 0.004 Hz (see panel {\it a} of Fig.~\ref{LF-Powspec}; Table 1). Only one power spectrum in hard state shows a significant ($1-3.28 \times 10^{-10} $) low-frequency quasi-periodic oscillation (LFQPO) at $0.452 \pm 0.0049$ Hz with a quality (Q) factor of $6.3 \pm 1.93$ (see Table 1 and panel {\it b} of Fig.~\ref{LF-Powspec}). A high intensity state (phase 5--7) power spectrum typically shows a very-low-frequency-noise (VLFN) below $\sim 1$ Hz with an RMS $\sim 6 - 10 \%$  having the Leahy-power reaching a few times 100 at 0.004 Hz. Such a power spectrum also shows a weak broad hump near 0.3 Hz (panel {\it c} of Fig.~\ref{LF-Powspec}; Table 1). A phase 8 power spectrum shows a red noise (RMS $\sim 5 - 25 \%$) roughly below 0.1 Hz with the Leahy-power reaching about $100$ at 0.004 Hz (panel {\it d} of Fig.~\ref{LF-Powspec}). In most cases, such a power spectrum has a significant peaked noise at $\sim 10$ Hz (see Table 1 and panel {\it e} of Fig.~\ref{LF-Powspec}). For example, the peaked noise in the data set of Nov 17, 2000 (03:46:10--04:45:07) has a significance of $1-7.9 \times 10^{-175}$, a centroid-frequency of $11.01 \pm 0.25$ Hz, an RMS-amplitude of $(5.8 \pm 0.14)\%$ and a Q-factor of $0.73 \pm 0.046$. A typical phase 9 power spectrum has a red noise (RMS $\sim 4 - 12\%$) below $\sim 0.02$ Hz with the Leahy-power reaching a few times 10 at 0.004 Hz (panel {\it f} of Fig.~\ref{LF-Powspec}).

In Fig.~\ref{CCD-HID}, we have displayed the locations of the thermonuclear bursts, the LFQPO and the kilohertz (kHz) QPO \citep{MukherjeeBhattacharyya2011} 
in the CD and the HID. While the non-photospheric-radius-expansion (non-PRE) 
bursts occured in the hard state, the PRE bursts and the kHz QPO appeared in phase 8 
(Table 1).

\section{Discussion and Conclusions}\label{Discussion}

\begin{figure*}
\centering
\includegraphics*[width=\textwidth]{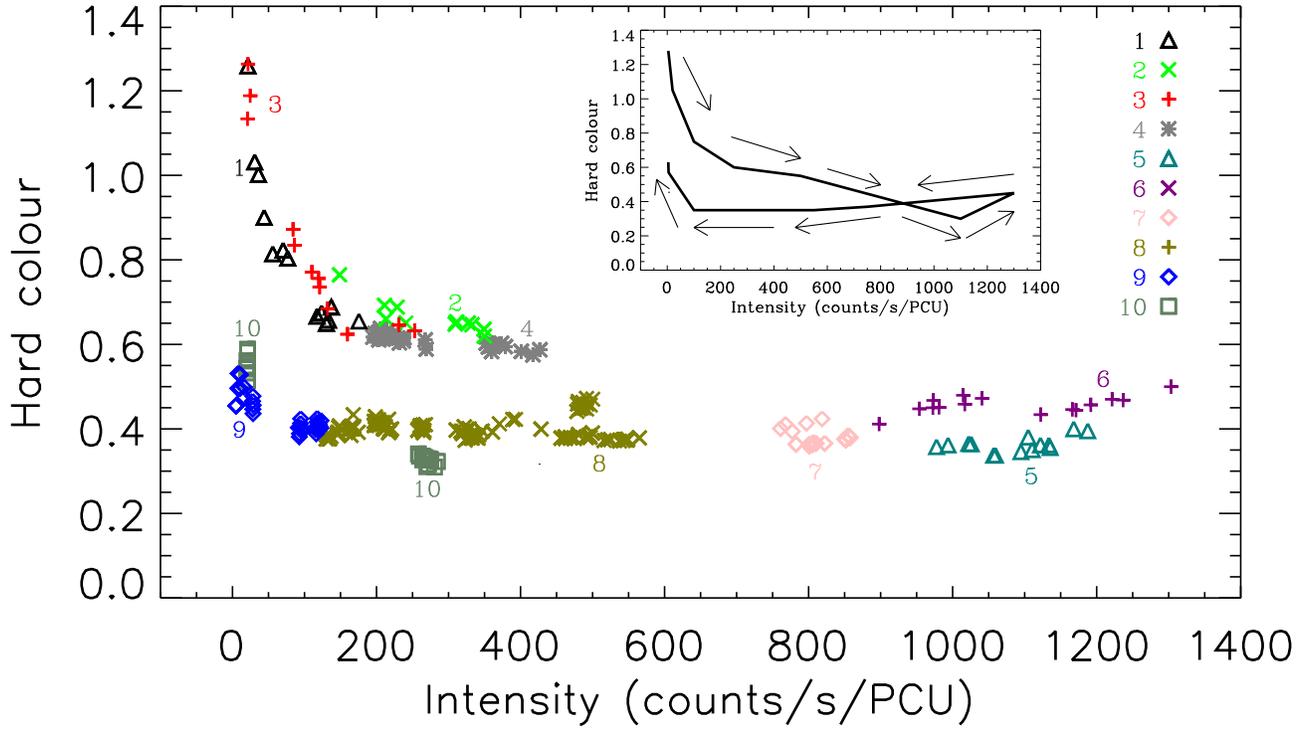}
\caption{Hardness-intensity diagram (HID) of EXO 1745--248 using the
{\it RXTE} PCA data. Hard colour and intensity (for PCU 2) are defined 
in \S~\ref{DataAnalysisandResults}.  
Various temporal segments (phases; see Table 1 
and \S~\ref{DataAnalysisandResults}) are shown with different 
symbols and phase numbers (see Table 1). The schematic in the inset shows the 
motion of the source along the HID track with time. This figure clearly 
shows hysteresis in the spectral states.
\label{HID}}
\end{figure*}

\begin{figure*}
\centering
\includegraphics*[width=\textwidth]{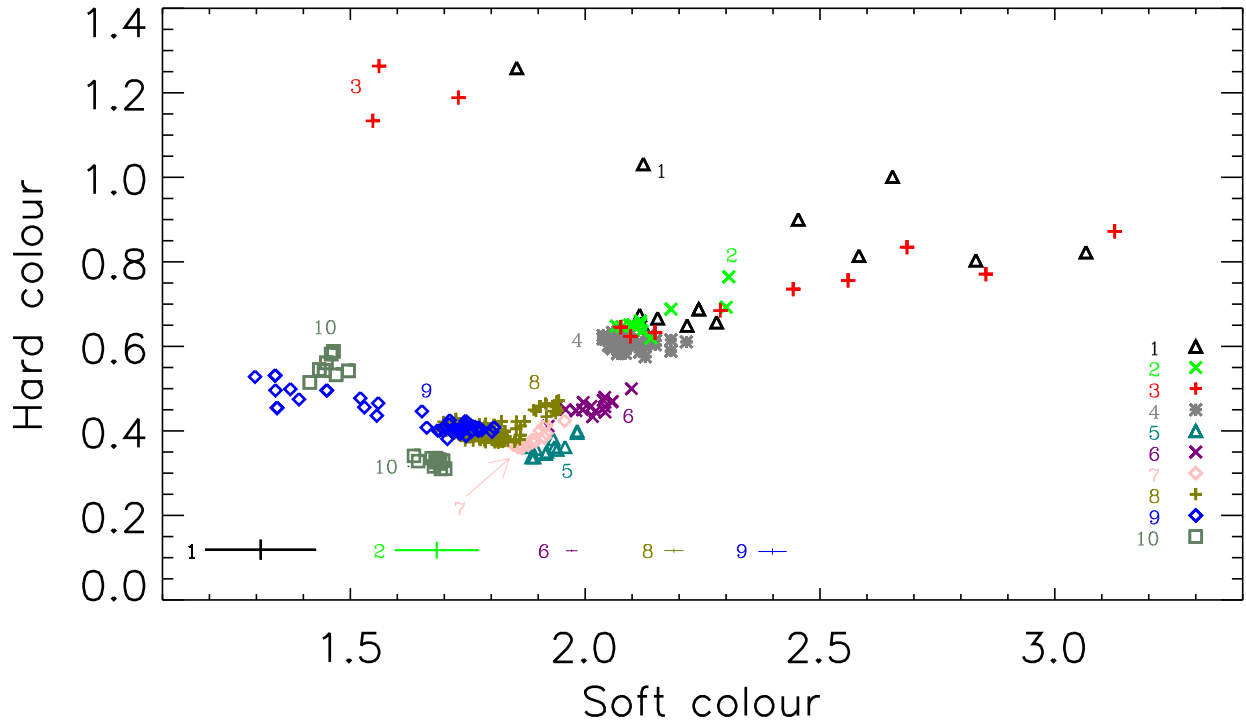}
\caption{Colour-colour diagram of EXO 1745--248 using the
{\it RXTE} PCA data. Hard colour and soft colour are defined 
in \S~\ref{DataAnalysisandResults}.
Various temporal segments (phases; see Table 1
and \S~\ref{DataAnalysisandResults}) are shown with different
symbols and phase numbers (see Table 1). Typical $1\sigma$ error
bars for some of the phases are shown.
This figure suggests hysteresis in the spectral states.
\label{CCD}}
\end{figure*}

\begin{figure*}
\centering
\begin{tabular}{c}
\hspace{-1.0cm}
\includegraphics*[width=\textwidth]{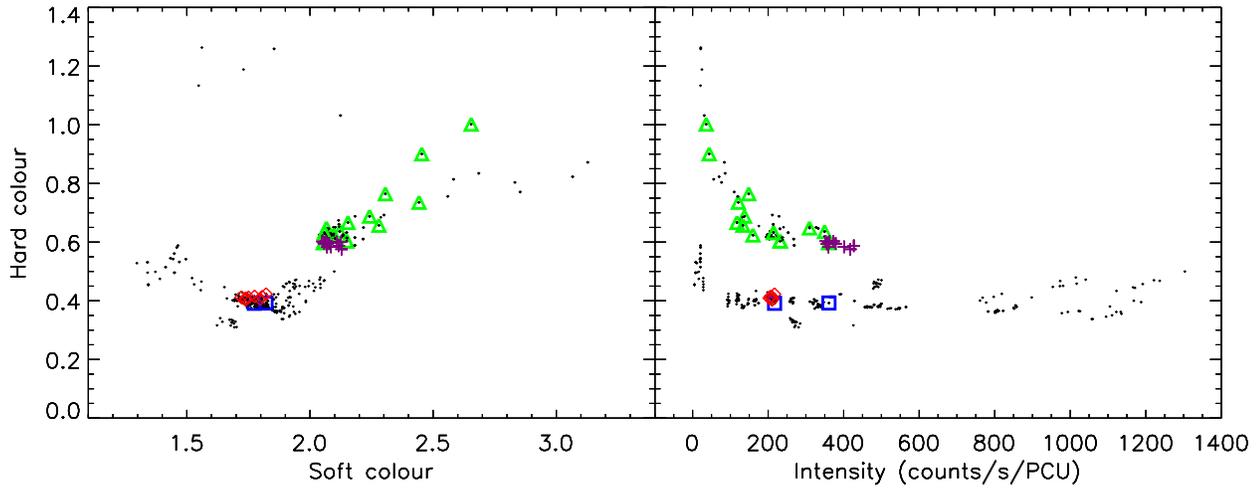}
\end{tabular}
\caption{Colour-colour diagram (left panel) and hardness-intensity diagram
of EXO 1745--248 (same as Fig.~\ref{CCD} and Fig.~\ref{HID}).
The triangles mark the non-PRE thremonuclear X-ray bursts and the two squares mark 
the PRE bursts (\S~\ref{DataAnalysisandResults}).
The plus signs mark the continuous data set containing the LFQPO 
and the diamond signs mark 
the continuous data set during which the kHz QPO appeared 
(\S~\ref{DataAnalysisandResults}).
\label{CCD-HID}}
\end{figure*}

In this Letter, we have studied the evolution of spectral states of the
neutron star LMXB EXO 1745--248. We tentatively conclude that it is an atoll source because 
of the following reasons. 
(1) From the spectral fitting, we find that the observed 
$2-30$ keV unabsorbed source flux varied in the range 
$(0.05-2.12)\times10^{-8}$ ergs cm$^{-2}$ s$^{-1}$. Such a large
intensity variation does not happen in a source, which shows an exclusive
`Z' behaviour \citep{vanderKlis2006}.
(2) The hard colours (Fig.~\ref{HID}) of EXO 1745--248 are consistent with
those of atoll sources, but different from Z sources \citep{Munoetal2002}. 
(3) The source shows parallel tracks for the higher hard colour values in HID
(Fig.~\ref{HID}), which are typical of atoll sources \citep{vanderKlis2006}.
(4) Shape of the CD track for lower hard colour values looks like a banana (Fig.~\ref{CCD}).
(5) Hard state to soft state transition of the source was plausibly quick \citep{vanderKlis2006}.
(6) PRE bursts were found in the softer state (Fig.~\ref{CCD-HID}), as usually
observed for fast spinning neutron star LMXBs \citep{Munoetal2004}. 
(7) The kHz QPO was observed in the transitional state (plausibly LB/LLB),
which is usual for atolls \citep{MaitraBailyn2004, vanderKlis2006}.
(8) VLFN at $< 1$ HZ was observed in BS, and $\sim 10$ Hz peaked noise was detected 
in the transitional state (plausibly LB/LLB), which are usual for atolls 
\citep{MaitraBailyn2004, vanderKlis2006}.
(9) Red noise RMS is higher in the hard state \citep{vanderKlis2006}.
However, although we cannot confirm, there is some chance that at the most 
intense state, the source transformed into a Z source (e.g., \citet{Homanetal2010}).
This is because, the estimated source luminosity in this state was
$\sim 0.5$ times the Eddington luminosity \citep{vanderKlis2006}, for a $2-30$ keV flux of 
$2.12\times10^{-8}$ ergs cm$^{-2}$ s$^{-1}$, and assuming a 5.5 kpc source
distance, 1.4 M$_\odot$ neutron star mass, 6.0 stellar radius-to-mass ratio 
and ionized hydrogenic accreted matter.

\begin{figure*}
\centering
\begin{tabular}{c}
\hspace{-1.0cm}
\includegraphics*[width=\textwidth]{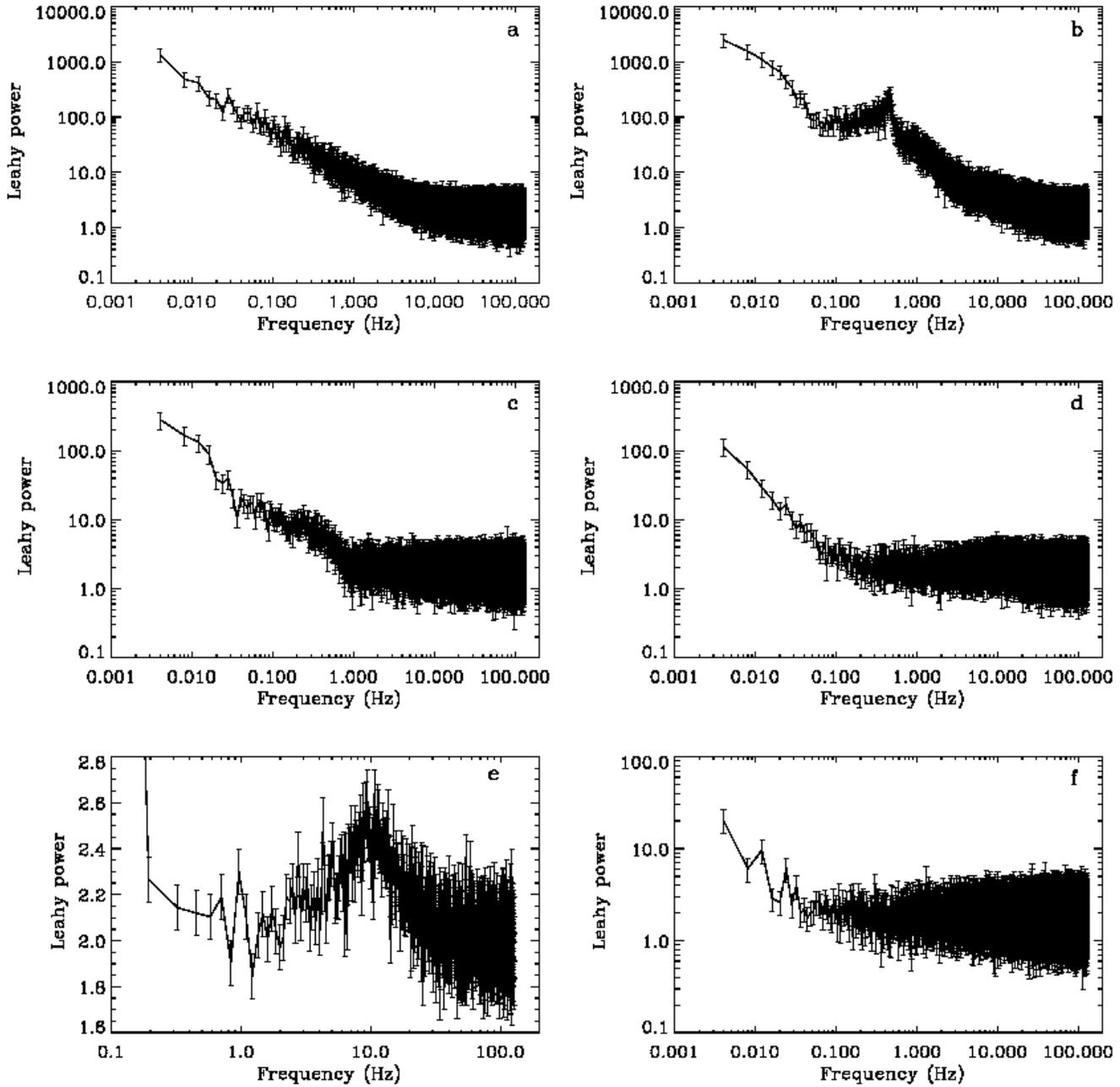}
\end{tabular}
\caption{Typical low-frequency power spectra of various phases of EXO 1745--248 (see Table 1
and \S~\ref{DataAnalysisandResults}). Panel {\it a}: phase 1--3.; panel {\it b}:
the power spectrum of phase 4 with LFQPO; panel {\it c}: phase 5--7;
panel {\it d}: phase 8; panel {\it e}: same as panel {\it d}, but in a different
scale to show the peaked noise clearly; and panel {\it f}: phase 9.
\label{LF-Powspec}}
\end{figure*}

EXO 1745--248 is very interesting, unusual and important for the following reasons. 
(1) The source exhibited a clear overall clockwise hysteresis in HID and CD 
(Fig.~\ref{HID} and \ref{CCD}). A local anti-clockwise hysteresis is also observed
in the high intensity state (plausibly UB).
(2) In the hard state (plausibly EIS), unlike a typical atoll, the hard colour changed 
largely, and no horizontal track is present at the highest hard colour in HID \citep{vanderKlis2006}.
Moreover, the hard-to-soft transition involved a large change in intensity, unlike
several black hole sources. These caused an HID-track-shape intermediate between
atoll `C' tracks and black hole `q' tracks.
(3) In CD/HID, the source moved from EIS to UB, while usually an atoll moves
to LB/LLB from EIS \citep{vanStraatenetal2003, vanderKlis2006}.
(4) The CD/HID tracks of EXO 1745--248 could be segmented in time 
(Fig.~\ref{HID} and \ref{CCD}). Several segments can be distinguished by timing
properties (Fig.~\ref{LF-Powspec}), 
which shows that these segments are actually in different states,
i.e., not in the same state shifted by secular motions. 
The source typically dwells in a segment for a few days to about a month
(Table 1). 
(5) EXO 1745--248 took at least 2.5 months to trace the BS, as opposed to a few
hours to a day, which is typical for an atoll source \citep{vanderKlis2006}.

The CD/HID hysteresis tracks of EXO 1745--248 could be very useful to relate the
accretion processes in neutron star systems and black hole systems (\S~\ref{Introduction}).
Finally, the HID hysteresis track of EXO 1745--248, which is
intermediate between the `q'-like hysteresis track of Aql X-1 and `C'-like
non-hysteresis tracks of most atoll sources, suggests that the popular
EIS-IS-BS framework of `C'-like tracks might be a special case of a more
general hysteresis behaviour. However, observations of more such intermediate
sources are required to verify this.

\clearpage
\begin{table*}
 \centering
\caption{Various properties of the source in the 10 phases (see
\S~\ref{DataAnalysisandResults}; see also Figs.~\ref{HID} and \ref{CCD}).}
\begin{tabular}{|c|c|c|c|c|c|p{5cm}|}

\hline 
Phase & Start time$^{1}$ & End time$^{2}$ & Soft-colour$^{3}$ & Hard-colour$^{4}$ & Intensity$^{5}$ & Remarks$^{6}$\tabularnewline
\hline
\hline 
Phase1 & 13/07/2000 04:43:28 & 13/07/2000 05:13:04 & 2.2418623 & 0.68810917 & 137.27226 & No narrow feature.\tabularnewline
\cline{2-6} 
 & 21/07/2000 10:23:28 & 21/07/2000 11:44:00 & 2.2798728 & 0.65725367 & 133.26667 & The Leahy-powers start rising\tabularnewline
\cline{1-6} 
Phase2 & 24/07/2000 15:15:28 & 24/07/2000 16:16:00 & 2.1178232 & 0.66027898 & 213.43736 & significantly above 2 in $\sim 2-12$ Hz,\tabularnewline
\cline{2-6} 
 & 06/08/2000 12:55:28 & 06/08/2000 13:28:00 & 2.0948016 & 0.65078129 & 240.75461 & and reach above 1000 at 0.004 Hz\tabularnewline
\cline{1-6} 
Phase3 & 06/08/2000 14:00:32 & 06/08/2000 15:10:08 & 2.1486752 & 0.63243319 & 252.97392 & (see Fig.~\ref{LF-Powspec}-a).\tabularnewline
\cline{2-6} 
 & 13/08/2000 09:58:24 & 13/08/2000 11:08:00 & 3.1275264 & 0.87214917 & 84.047352 & Non-PRE bursts observed.\tabularnewline
\hline 
Phase4 & 13/08/2000 11:39:28 & 13/08/2000 12:50:08 & 2.0437827 & 0.61713214 & 195.17507 & 
Similar to the phases $1-3$. An LFQPO\tabularnewline
\cline{2-6} 
 & 15/08/2000 17:57:52 & 15/08/2000 19:26:40 & 2.054908 & 0.59863458 & 360.6038 & is detected in one segment (Fig.~\ref{LF-Powspec}-b).\tabularnewline
\hline 
Phase5 & 18/08/2000 13:08:32 & 18/08/2000 13:41:36 & 1.9404588 & 0.35567838 & 1135.4601 & The Leahy-powers start rising\tabularnewline
\cline{2-6} 
 & 21/08/2000 19:36:32 & 21/08/2000 19:55:44 & 1.8951522 & 0.35778984 & 977.661 & significantly above 2 below $\sim 1$ Hz,\tabularnewline
\cline{1-6} 
Phase6 & 24/08/2000 10:48:48 & 24/08/2000 11:19:44 & 2.0420255 & 0.45687916 & 1191.8152 & 
and reach above 100 at 0.004 Hz.\tabularnewline
\cline{2-6} 
 & 27/08/2000 05:51:28 & 27/08/2000 06:28:00 & 1.9212794 & 0.41142752 & 897.9504 & A plausible hump at $\sim0.3$ Hz is\tabularnewline
\cline{1-6} 
Phase7 & 27/08/2000 06:53:36 & 27/08/2000 08:10:56 & 1.8786671 & 0.36700101 & 806.13514 & typically seen (see Fig.~\ref{LF-Powspec}-c).\tabularnewline
\cline{2-6} 
 & 30/08/2000 15:23:28 & 30/08/2000 15:46:40 & 1.9175748 & 0.39400707 & 777.30473 & \tabularnewline
\hline 
Phase8 & 05/09/2000 09:41:20 & 05/09/2000 11:13:04 & 1.809879 & 0.37902471 & 340.92112 & The Leahy-powers start rising\tabularnewline
\cline{2-6} 
 & 06/10/2000 05:00:32 & 06/10/2000 05:13:36 & 1.75375 & 0.38471589 & 164.56586 & significantly above 2 below $\sim 0.1$ Hz, and reach $\approx 100$ at 0.004 Hz. A peaked noise around 10 Hz is typically seen (see Figs.~\ref{LF-Powspec}-d and \ref{LF-Powspec}-e). PRE bursts and kHz QPO found in this phase.\tabularnewline
\hline 
Phase9 & 09/10/2000 05:13:20 & 09/10/2000 05:50:56 & 1.732144 & 0.4048109 & 120.68564 & The Leahy-powers start rising\tabularnewline
\cline{2-6} 
 & 03/11/2000 00:02:24 & 03/11/2000 00:17:52 & 1.3439725 & 0.45464597 & 4.9097872 & significantly above 2 below $\sim 0.02$ Hz, and reach above 10 at 0.004 Hz (Fig.~\ref{LF-Powspec}-f).\tabularnewline
\hline 
Phase10 & 02/07/2002 20:38:24 & 02/07/2002 20:47:44 & 1.6929414 & 0.30983275 & 280.54887 & Roughly similar to phase 9.\tabularnewline
\cline{2-6} 
 & 22/07/2002 08:46:24 & 22/07/2002 11:04:00 & 1.4488735 & 0.56095016 & 20.41566 & \tabularnewline
\hline
\end{tabular}
\begin{flushleft}
$^1$Start time of the first (upper line) and the last (lower line) continuous time segments of the phase.\\
$^2$ End time of the first (upper line) and the last (lower line) continuous time segments of
 the phase.\\
$^3$Soft-colours (defined in \S~\ref{DataAnalysisandResults}) of the first time bin of the first continuous time segment (upper line), and the last time bin of the last continuous time segment (lower line) of the phase.\\
$^4$Hard-colours (defined in \S~\ref{DataAnalysisandResults}) of the first time bin of the first continuous time segment (upper line), and the last time bin of the last continuous time segment (lower line) of the phase.\\ 
$^5$Intensities (defined in \S~\ref{DataAnalysisandResults}) of the first time bin of the first continuous time segment (upper line), and the last time bin of the last continuous time segment (lower line) of the phase.\\
$^6$Primarily a short description of a typical power spectrum of the phase (\S~\ref{DataAnalysisandResults}).\\
\end{flushleft}
\end{table*}

\clearpage

\clearpage

\clearpage

\clearpage

\end{document}